# Transport of platelets induced by red blood cells based on mixture theory

Wei-Tao Wu[1], Nadine Aubry[2], James F. Antaki[1], Mehrdad Massoudi[3*]

1. Department of Biomedical Engineering, Carnegie Mellon University, Pittsburgh, PA, 15213, USA

2. Department of Mechanical and Industrial Engineering, Northeastern University, Boston, MA 02115, USA

3. U. S. Department of Energy, National Energy Technology Laboratory (NETL), Pittsburgh, PA, 15236, USA

Address for correspondence:

Mehrdad Massoudi, PhD,

U. S. Department of Energy,

National Energy Technology Laboratory (NETL),

626 Cochrans Mill Road, P.O. Box 10940,

Pittsburgh, PA. 15236.

Email: Mehrdad.Massoudi@NETL.DOE.GOV

**Abstract**

Near-wall enrichment of platelets strongly influences thrombus formation *in vivo* and *in vitro*. This paper develops a multi-constituent continuum approach to study this phenomenon. A mixture-theory model is used to describe motion of plasma and red blood cells (RBCs) and interactions between the two components. A transport model is developed to study influence of the RBCs field on the platelets. The model is used to study blood flow in a rectangular micro-channel, a sudden expansion micro-channel, and a channel containing micro crevices (representing a practical problem encountered in most blood-wetted devices). The simulations show that in the rectangular channel the concentration of the platelets near the walls is about five times higher than the concentration near the channel centerline. It is also noticed that in the channel with crevices, extremely a large number of platelets accumulate in the deep part of the crevices and this may serve as the nidus for excessive thrombus formation occurring in medical devices.

## 1. Introduction

A thrombus, sometimes called a blood clot, is a natural response of the body to an injury by attempting to prevent bleeding; however, excessive thrombosis is responsible for diseases such as stroke and device malfunction. Thrombosis in the coronary artery can lead to heart attacks; a thrombus can also be transported to the brain by blood circulation causing cardiogenic strokes (Furie & Furie, 2008; Handin, 2005). Thrombus generation in blood-related medical devices can reduce the efficiency of the devices and this can lead to their malfunction (Jaffer, Fredenburgh, Hirsh, & Weitz, 2015; Kirklin et al., 2012; Reviakine et al., 2016; Slaughter et al., 2009).

Thrombosis is usually initiated as platelet adhesion on biological or artificial surfaces. Thrombus initiation and growth in vessels or on medical devices is strongly influenced by the number density of the platelets near the walls and surfaces (Cito, Mazzeo, & Badimon, 2013; Skorczewski, Erickson, & Fogelson, 2013; F. F. Weller, 2008; Yang, Jäger, Neuss-Radu, & Richter, 2016). In small blood vessels, the platelets tend to move away from the center of the blood vessels and accumulate near the walls. Such a transverse migration of platelets is believed to be closely related to the motion of the red blood cells (RBCs), where opposite to the movement of the platelets towards the walls, the RBCs tend to move away from the walls of the vessels (AlMomani, Udaykumar, Marshall, & Chandran, 2008; Goldsmith & Turitto, 1986; Vincent T. Turitto, Benis, & Leonard, 1972). Through many mesoscale simulations, the non-uniform distribution of the platelets is believed to be mainly attributed to the collision between the RBCs and the platelets (AlMomani et al., 2008; Reasor, Mehrabadi, Ku, & Aidun, 2013; Skorczewski et al., 2013). In mesoscale simulations, the microstructure of the blood is modeled by computationally predicting the dynamics of each individual blood cell and the fluid (plasma) flow. Direct experimental measurements also indicate that when the RBCs are introduced in the flow, the platelets migration to the vessel walls increases intensively (AlMomani et al., 2008; Cadroy & Hanson, 1990; Joist, Bauman, & Sutera, 1998; Peerschke et al., 2007; V T Turitto & Weiss, 1980).

In order to describe or to be able to predict the non-uniform distribution of the platelets, we need to use a multi-component theory to model blood flow so that important quantities such as the velocity and the volume fraction fields of the RBCs can be studied. That is, a single-phase blood model is not able to predict the migration of the blood cells. Although many numerical reproductions of the near-wall platelets enrichment have been done using mesoscale simulations, in most engineering-scale applications, due to the high computational cost of the mesoscale simulations, a continuum model has to be applied. In the present study, using the multi-component blood model recently developed by (Wu, Aubry, Massoudi, Kim, & Antaki, 2014; Wu, Yang, Antaki, Aubry, & Massoudi, 2015), we introduce and advocate a transport flux equation to study and model the non-uniform distribution of the platelets .

## 2. Methods: The Mathematical Model

A thrombus usually indicates a blood clot attached to the (damaged) vascular walls. There are a few theoretical models of thrombus formation [see (M. Anand, Rajagopal, & Rajagopal, 2006a, 2006b; M Anand & Rajagopal, 2004; M Anand, Rajagopal, & Rajagopal, 2008; Mohan Anand & Rajagopal, 2002; Flamm & Diamond, 2012; Kuharsky & Fogelson, 2001; Wu et al., 2017)]. In general, it is believed that the RBCs distribution can influence the distribution of the platelets which in turn is closely related to the thrombus formation (M Anand et al., 2008). In this paper, we assume that blood is a multi-component fluid composed of red blood cells, plasma and platelets. The presence and influence of other components such as the white bloods cells are ignored. Furthermore, we assume that the motion of the RBCs and the plasma are governed by the conservation equations based on mixture theory, whereas for the motion of the platelets we propose a transport flux (a convection-diffusion) equation. These conservation equations are used to obtain the velocity fields for the plasma and the RBCs, the hematocrit (RBCs concentration), and the pressure fields. The equations of motion used in this paper are based on the Mixture Theory (theory of interacting continua) as given in (Massoudi, Kim, & Antaki, 2012; Rajagopal & Tao, 1995; Wu, Aubry, Massoudi, et al., 2014). The classical theory of mixtures is described in detail in books by Truesdell (1984) (Truesdell, 1984), Rajagopal and Tao (1995) (Rajagopal & Tao, 1995), and in review articles by Bowen (1976) (Bowen, 1976) and Atkin and Craine (1976 a,b) (Atkin & Craine, 1976a, 1976b).

## 2.1 Governing equations

### 2.1.1.Conservation of mass

In the absence of thermo-chemical and electromagnetic effects, the governing equations consist of the conservation of mass, linear momentum and angular momentum. In the Eulerian form, conservation of mass, for each component, is expressed as (Bowen, 1976):

$$\frac{\partial \rho_f}{\partial t} + div(\rho_f \boldsymbol{v}_f) = 0 \tag{1}$$

$$\frac{\partial \rho_s}{\partial t} + div(\rho_s \boldsymbol{v}_s) = 0 \tag{2}$$

where $\frac{\partial}{\partial t}$ is the derivative with respect to time, *div* is the divergence operator and $\boldsymbol{v}$ is the velocity field. The subscript 'f' refers to the fluid (plasma), and 's' to the solid particles, representing the RBCs. The densities of the two constituents are: $\rho_f = (1 - \phi)\rho_{f0}$, $\rho_s = \phi \rho_{s0}$, where $\rho_{f0}$ and $\rho_{s0}$ are the pure density of the plasma and the RBCs in the reference configuration, respectively, and $\phi$ is the volume fraction (hematocrit) of the RBCs. $\rho_{f0}$ and $\rho_{s0}$ are constant in this paper.

### 2.1.2. Conservation of linear momentum

The balance of the linear momentum can be written as:

$$\rho_f \frac{D\boldsymbol{v}_f}{Dt} = div\boldsymbol{T}_f + \rho_f \boldsymbol{b}_f + \boldsymbol{f}_I \tag{3}$$

$$\rho_s \frac{D\boldsymbol{v}_s}{Dt} = div\boldsymbol{T}_s + \rho_s \boldsymbol{b}_s - \boldsymbol{f}_I \tag{4}$$

where, $\frac{D}{Dt}$ is the material derivative. For any scalar $\beta$, $\frac{D\beta}{Dt} = \frac{\partial \beta}{\partial t} + \boldsymbol{v} \cdot grad\beta$; for any vector $\boldsymbol{w}$, $\frac{D\boldsymbol{w}}{Dt} = \frac{\partial \boldsymbol{w}}{\partial t} + (grad\boldsymbol{w})\boldsymbol{v}$, where 'grad' is the gradient operator, $\boldsymbol{T}_f$ and $\boldsymbol{T}_s$ are the partial Cauchy stress tensors for the plasma and the RBCs, respectively. $\boldsymbol{f}_I$ represents the interaction force (exchange of momentum) between the two components, and $\boldsymbol{b}_f$ and $\boldsymbol{b}_s$ refer to body forces. The balance of the angular momentum implies that, in the absence of couple stresses, the total Cauchy stress tensor is symmetric. To close these equations, constitutive relations are needed for the stress tensors $\boldsymbol{T}_f$ and $\boldsymbol{T}_s$ and the interaction force $\boldsymbol{f}_I$.

### 2.1.3. Conservation of platelets concentration

In this paper, we do not consider blood as a three-component fluid (plasma, RBCs and platelets). However, we recognize that the platelets move and are deformed by the flow. While we ignore the effects of the platelets on the plasma and the RBCs, in order to describe the motion of the platelets, we propose a convection-diffusion equation of the type(Bridges, Karra, & Rajagopal, 2010; Massoudi & Uguz, 2012) due to the motion of the mixture, composed of plasma and RBCs:

$$\frac{\partial C}{\partial t} + div\boldsymbol{v}_m C = div\boldsymbol{Q} \tag{5}$$

where $\boldsymbol{v}_m = \phi\boldsymbol{v}_s + (1 - \phi)\boldsymbol{v}_f$ is the velocity of the whole blood flow, $C$ is the concentration of the platelets; $\boldsymbol{Q}$ refers to the diffusion flux of the platelets in blood (Hund & Antaki, 2009). In the next section, we discuss the constitutive relations used in this paper.

## 2.2. Constitutive relations

As mentioned earlier, blood is a complex non-linear fluid, composed of red blood cells (RBCs), white blood cells (WBCs), platelets, plasma. Under normal conditions, the volume fraction (hematocrit) of the RBCs is about 45%, and as a result, the rheological properties of the (whole) blood is greatly influenced by the distribution of the RBCs. Among the unusual observations encountered in blood flow, one can name the aggregation and the disaggregation of the RBCs, related to the shear rate, the degree of their deformability, and their alignment responding to extensional flows [see Robertson et al. (2008) (Robertson, Sequeira, & Kameneva, 2008), Popel and Johnson (2005) (Popel & Johnson, 2005), Bäumler et al.(1999) (Bäumler, Neu, Donath, & Kiesewetter, 1999), Chien (1970) (Chien, 1970), Wu et al. (2014) (Wu, Aubry, Massoudi, et al., 2014)]. In order to capture some or all of these effects on a macroscopic level, one can model blood, for example, as a shear-thinning fluid with stress relaxation (Bagchi, 2007). For micro-scale applications, for example blood flow in a vessel with the diameter in the range of 20 to 500 microns (and shear rates below 100 s$^{-1}$), other non-linear or unusual phenomena such as their non-linear distribution of volume fraction can occur (Kameneva, Garrett, Watach, & Borovetz, 1998; Middleman, 1972; Rourke & Ernstene, 1930). The viscoelastic behavior of blood has been reported by Thurston (G B Thurston, 1972; George B Thurston, 1973) and others. It is known that when the shear rate is low the stress relaxation becomes more significant. To model the RBCs we use and modify the model proposed by Yeleswarapu et. al (K K Yeleswarapu, Kameneva, Rajagopal, & Antaki, 1998; Krishna K Yeleswarapu, Antaki, Kameneva, & Rajagopal, 1995), which captures the shear-thinning behavior of blood over a wide range of shear rates; it is a generalization of a three constant Oldroyd-B fluid. Therefore in some sense, the model that we are using for the RBCs (and blood as a whole) is a viscoelastic shear-thinning fluid model.

### 2.2.1. Stress tensor of the plasma

We assume that the plasma behaves as a viscous fluid,

$$\boldsymbol{T}_f = \left[-p(1-\phi) + \lambda_f(1-\phi)tr\boldsymbol{D}_f\right]\boldsymbol{I} + 2\mu_f(1-\phi)\boldsymbol{D}_f \tag{6}$$

where $p$ is the pressure of the mixture, $\lambda_f$ and $\mu_f$ are the (constant) first and the second coefficients of viscosity of the pure plasma, $\boldsymbol{D}_f = \frac{1}{2}\left[\left(grad\ \boldsymbol{v}_f\right) + \left(grad\ \boldsymbol{v}_f\right)^T\right]$, '$tr$' stands for the trace of a second order tensor, and $\boldsymbol{I}$ is the identity tensor. [See (Wu, Aubry, Massoudi, et al., 2014) for further details and derivation of these equations].

### 2.2.2. Stress tensor of the RBCs

The RBCs are assumed to behave as a viscoelastic shear-thinning fluid, whose viscosity, $\mu_s$ is a function of the volume fraction, $\phi$:

$$\boldsymbol{T}_s = \left[-p\phi + \beta_{20}(\phi + \phi^2)tr\boldsymbol{D}_s\right]\boldsymbol{I} + \mu_s(\phi, tr\boldsymbol{D}_s^2)\phi\boldsymbol{D}_s \tag{7}$$

$$\mu_s(\phi, tr\boldsymbol{D}_s^2)\phi = \left(\mu_\infty(\phi) + \left(\mu_0(\phi) - \mu_\infty(\phi)\right)\frac{1 + ln\left(1 + k(\phi)(2tr\boldsymbol{D}_s^2)^{1/2}\right)}{1 + k(\phi)(2tr\boldsymbol{D}_s^2)^{1/2}}\right) \tag{8}$$

Where $\boldsymbol{D}_s = \frac{1}{2}[(grad\ \boldsymbol{v}_s) + (grad\ \boldsymbol{v}_s)^T]$ and $\beta_{20}$ is (similar to) the second coefficient of viscosity .These equations are calibrated using the experimental measurements of Brooks et al., over a range of hematocrit from 8.6%-70.2%: (Brooks, Goodwin, & Seaman, 1970) (Also see Figure 1):

$$\mu_0 = 537.002\phi^3 + 55.006\phi^2 - 0.129\phi \tag{9}a$$

$$\mu_\infty = 27.873\phi^3 - 21.218\phi^2 + 14.439\phi \tag{9}b$$

$$k = 11 \tag{9}c$$

The above equations account for low volume fraction of RBCs that occurs in blood flow at micro-scale (Wu et al., 2015). It is noted that equations (9)a and (9)b imply that the viscosity of the RBCs approaches zero when the hematocrit approaches zero.

### 2.2.3.Interaction forces between plasma and RBCs

For the interaction force between the two components, we only consider terms which correspond to the Stokes drag force and the Saffman's shear- lift force[See (Johnson, Massoudi, & Rajagopal, 1991; Massoudi, 2003)],

$$\boldsymbol{f}_I = \frac{9\mu_f}{2a^2}\phi f(\phi)\left(\boldsymbol{v}_s - \boldsymbol{v}_f\right) + \frac{3(6.46)\left(\rho_f\mu_f\right)^{1/2}}{4\pi a}\phi\left(2tr\boldsymbol{D}_f{}^2\right)^{-1/4}\boldsymbol{D}_f\left(\boldsymbol{v}_s - \boldsymbol{v}_f\right) \tag{10}$$

where $a$ is the (hydraulic) diameter of the RBCs, here assumed to be $8\mu m$, and $f(\phi) = \exp(2.68\phi) + \phi^{0.43}$ is the the hindrance (drag) function suggested by Rusche and Issa (Rusche & Issa, 2000). For more discussions on the interaction force, see (Wu, Aubry, & Massoudi, 2014) and (Massoudi & Antaki, 2008).

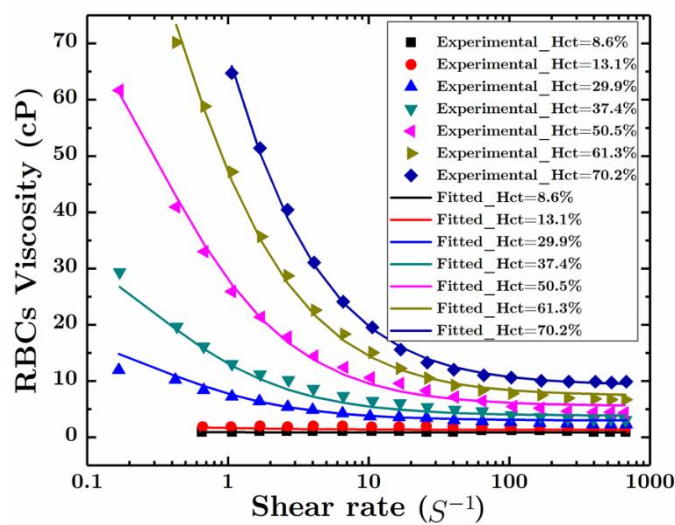

Figure 1. Viscosity of the RBCs as a function of the shear rate. Experimental data is by Brooks (Brooks et al., 1970).

2.2.4. The diffusive flux of the platelets

In this part of the paper, we discuss the modeling of the platelets. Usually the diffusivity of a species transported in flow can be described using the Brownian diffusion approach. However, in micro-scale flows, as many simulations and experiments have revealed, due to the collisions between the RBCs and the platelets, the platelets tend to move to the vessel walls. As a result the near-wall concentration of the platelets can be several times higher than near the vessel center (AlMomani et al., 2008; Reasor et al., 2013; Rui Zhao et al., 2008). In order to model the collision-induced platelets diffusion, based on the ideas proposed by Phillips et al. (Phillips, Armstrong, Brown, Graham, & Abbott, 1992), Wooton et. al (Wootton & Ku, 1999) and Hund & Antaki (Hund & Antaki, 2009), we assume that the platelets diffusion flux, $\boldsymbol{Q}$, is given by ,

$$\boldsymbol{Q} = \boldsymbol{Q}_1 + \boldsymbol{Q}_2 + \boldsymbol{Q}_3 \tag{11}$$

where

$$\boldsymbol{Q}_1 = D_C \nabla C \tag{12}$$

$$\boldsymbol{Q}_2 = D_C q(\phi) \nabla C \tag{13}$$

$$\boldsymbol{Q}_3 = D_C C \nabla \big( q(\phi) \big) \tag{14}$$

$$D_C = (D_B + \xi f_s(\phi) \dot{\gamma}_m) \tag{15}$$

Where $\nabla$ is the 'del' operator. The term $\boldsymbol{Q}_1$ represents the diffusive flux due to the gradient of the platelet concentration, $\boldsymbol{Q}_2$, is the RBCs-enhanced diffusion of the platelets, and $\boldsymbol{Q}_3$ is the diffusive flux of the platelets responsible for the movement of the platelets away from the regions of high concentration of the RBCs. $\dot{\gamma}_m = \big( 2 tr \boldsymbol{D}_m{}^2 \big)^{1/2}$ is the shear rate of the mixture (whole) blood, $q(\phi)$ (to be discussed below) is a function of $\phi$, based on the experimental data, $D_B = 1.58 \times 10^{-13} m^2 s^{-1}$ is the Brownian diffusion constant, and $C$ represents the concentration of the platelets. The factor $D_C = (D_B + \xi f_s(\phi) \dot{\gamma}_m)$ represents the strength or the dependence of the platelets diffusion flux on the local shear rate of the mixture, assumed proportional to the particles collision frequency. For the sake of simplicity, we assume $f_s(\phi) = \phi$; in general, this function reduces to the Brownian diffusion constant as $\phi = 0$. Also, $\xi = 6.0 \times 10^{-14} m^{-2}$ is an empirical constant based on experiments(Goodman, Barlow, Crapo, Mohammad, & Solen, 2005; Wu et al., 2017).

We assume that $q(\phi)$ has the following properties: (1) $q(\phi)$ is a monotonic increasing function of $\phi$, so that the platelets are always excluded by the RBCs and that they diffuse to a region with lower volume fraction of the RBCs; (2) When $\phi = 0$, $q(\phi) = 0$, which indicates that when there are no RBCs, the collision-induced diffusion effect disappears. In this paper, we simply represent $q(\phi)$ as a polynomial function, where,

$$q(\phi) = \sum_{i}^{N} A_i \phi^i; i \geq 1, A_i \geq 0. \tag{16}$$

The specific form of this function will be discussed in Section 3 and we will see that for our paper, Equation 16, based on the available experimental data can be given by a second order polynomial.

The velocity, the volume fraction and the platelets concentration fields are, in general, given by:

$$\begin{cases} \boldsymbol{v}_f = v_{fx}(x,y,z;t)\boldsymbol{e_x} + v_{fy}(x,y,z;t)\boldsymbol{e_y} + v_{fz}(x,y,z;t)\boldsymbol{e_z} \\ \boldsymbol{v}_s = v_{sx}(x,y,z;t)\boldsymbol{e_x} + v_{sy}(x,y,z;t)\boldsymbol{e_y} + v_{sz}(x,y,z;t)\boldsymbol{e_z} \\ \phi = \phi(x,y,z;t) \\ C = C(x,y,z;t) \end{cases} \tag{17}$$

Using equation (17), substituting equations (6) ~ (10) into equations (3) ~ (4), and equations (11) ~ (16) into equation (5), we obtain the momentum equations in the vectorial form (here we have assumed that both components are incompressible) and the transport equation for the platelets in the scalar form.

The momentum equation for plasma is:

$$(1 - \phi)\rho_f \left[ \frac{\partial \boldsymbol{v}_f}{\partial t} + (grad \ \boldsymbol{v}_f)\boldsymbol{v}_f \right]$$

$$= -grad\big((1-\phi) \ p\big) + div \ \big(2\mu_f(1-\phi)\boldsymbol{D}_f\big) + \rho_f(1-\phi)\boldsymbol{b}_f + \frac{9\mu_f}{2a^2}f(\phi)\big(\boldsymbol{v}_s - \boldsymbol{v}_f\big) \tag{18}$$

$$+ \frac{3(6.46)\big(\rho_f\mu_f\big)^{1/2}}{4\pi a}\phi\big(2tr\boldsymbol{D}_f{}^2\big)^{-1/4}\boldsymbol{D}_f\big(\boldsymbol{v}_s - \boldsymbol{v}_f\big)$$

And for the RBCs,

$$\phi\rho_s \left[ \frac{\partial \boldsymbol{v}_s}{\partial t} + (grad \ \boldsymbol{v}_s)\boldsymbol{v}_s \right]$$

$$= -grad(\phi p) + div \left( \left( \mu_\infty(\phi) + \big(\mu_0(\phi) - \mu_\infty(\phi)\big)\frac{1 + ln\left(1 + k\big(2tr\boldsymbol{D}_s{}^2\big)^{1/2}\right)}{1 + k\big(2tr\boldsymbol{D}_s{}^2\big)^{1/2}} \right) \boldsymbol{D}_s \right) \tag{19}$$

$$+ \rho_s\phi\boldsymbol{b}_s - \frac{9\mu_f}{2a^2}f(\phi)\big(\boldsymbol{v}_s - \boldsymbol{v}_f\big) - \frac{3(6.46)\big(\rho_f\mu_f\big)^{1/2}}{4\pi a}\phi\big(2tr\boldsymbol{D}_f{}^2\big)^{-1/4}\boldsymbol{D}_f\big(\boldsymbol{v}_s - \boldsymbol{v}_f\big)$$

The transport (convection-diffusion) equation for the platelets is,

$$\frac{\partial C}{\partial t} + div(\boldsymbol{v_m}C) = div\left((D_B + \xi\phi\dot{\gamma})\nabla\big(1 + q(\phi)\big)C\right) + div\left((D_B + \xi\phi\dot{\gamma})\big(1 + q(\phi)\big)\nabla C\right) \tag{20}$$

These equations are subject to the boundary conditions summarized in Table 1. Numerically we use the finite volume method to solve the mathematical model described above and a computational fluid dynamics (CFD) solver is developed using the libraries of OpenFOAM®. This is a C++ toolbox for the development of customized numerical solvers, and pre-/post-processing utilities for the solution of continuum mechanics problems, including CFD applications (OpenCFD, 2011). For the details of the numerical algorithms dealing with the two-fluid (Eulerian-Eulerian) approach, see (Kim, 2012; Rusche, 2002; H. G. Weller, 2002).

Table 1. The Boundary conditions. For more details see (Rusche, 2002; Wu, Aubry, Antaki, & Massoudi, 2016; Wu et al., 2015).

| Boundary | Pressure | Velocity | Volume Fraction | Platelets Concentration |
|----------|----------|----------|-----------------|-------------------------|
| Inlet | Fixed flux | Fixed value | Fixed value | Fixed value |
| Outlet | Fixed value (reference) | Zero gradient | Zero gradient | Zero gradient |
| Wall | Fixed flux | Fixed value (0) | No flux | No flux |

## 3. Results

In this section, we first discuss the procedure for obtaining the form of $q(\phi)$. After calibrating this equation, i.e., finding the values for the coefficients, by looking at the data in a rectangular micro-channel, we study the platelets enrichment problem in a sudden expansion micro-channel and a channel with micro-crevices. In the following cases, for each geometry, the domain is discretized as hexahedral meshes using ICEM (ICEM, 2012). In each case, mesh-dependence studies are performed. To save computational time and expense, for all the cases that we have considered, if possible we have taken advantage of the symmetry of the problem. We use the following values for the material properties: $\mu_f = 0.96$cP, $\rho_f = 1027$kg/m$^3$ and $\rho_s = 1093$kg/m$^3$.

### 3.1. Flow in a rectangular micro-channel

The function representing the RBCs-induced diffusion of the platelets, $q(\phi)$, is obtained using the experimental data in a rectangular micro-channel by (R. Zhao, Marhefka, Antaki, & Kameneva, 2010). Since we need to study the various parameters appearing in the model, and in order to reduce the computational cost, we assume that the flow is two-dimensional. Figure 2 (a) shows the schematic of the rectangular micro-channel. According to Zhao's experiments (R. Zhao et al., 2010), the inlet velocity and the hematocrit are 0.4956 m/s and 36% respectively. The platelets bulk concentration is $C_0 = 2.5 \times 10^{14}$PLTs/m$^3$. As a result, the best fitted function, for $q(\phi)$ is

$$q(\phi) = 0.26\phi + 10.6\phi^2 + 537\phi^3 \tag{21}$$

A comparison of the platelets concentration along the y-direction obtained from the experimental measurement and the numerical prediction using our proposed model (equations (11)-(15)), is shown in Figure 2 (b). The relative concentration of the platelets is defined as the local platelets concentration divided by the bulk concentration of the platelets: $C/C_0$. From Figure 2 (b), we can see that a good agreement is achieved. The enrichment of the platelets near the channel walls is clearly demonstrated by the numerical simulations. The platelets concentration near the wall is about 5 times higher than the concentration near the centerline of the channel. Figure 3 (a) and (b) shows the development of the platelets and the RBCs profile along the x-direction; we can see that the platelets and the RBCs distribution become more non-uniform as the we move in the x-coordinate, due to the RBCs-induced transport flux and the shear lift force respectively. Furthermore, the profile of the platelets concentration develops faster near the entrance, which may be attributed to the uniform inlet boundary condition; this phenomenon was also observed experimentally by Zhao et al. (2007) (Rui Zhao, Kameneva, & Antaki, 2007).

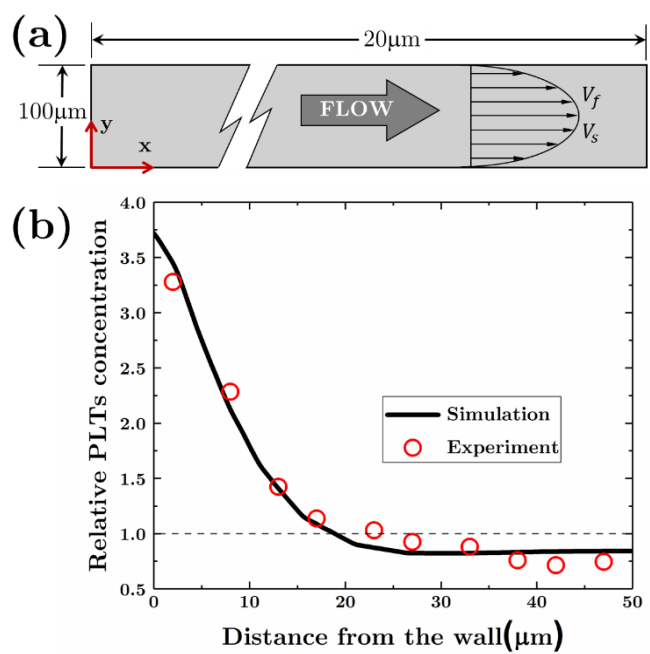

Figure 2 (a) Schematic of the rectangular micro-channel. The inlet velocity is 0.4956 m/s. The hematocrit is 36%. The platelets bulk concentration is $C_0 = 2.5 \times 10^{14} \text{PLTs/m}^3$. (b) Comparison of the numerical and the experimental values for the relative platelets concentration along the y-direction (R. Zhao et al., 2010). The relative concentration of the platelets is defined as the local concentration of the platelets divided by the bulk concentration.

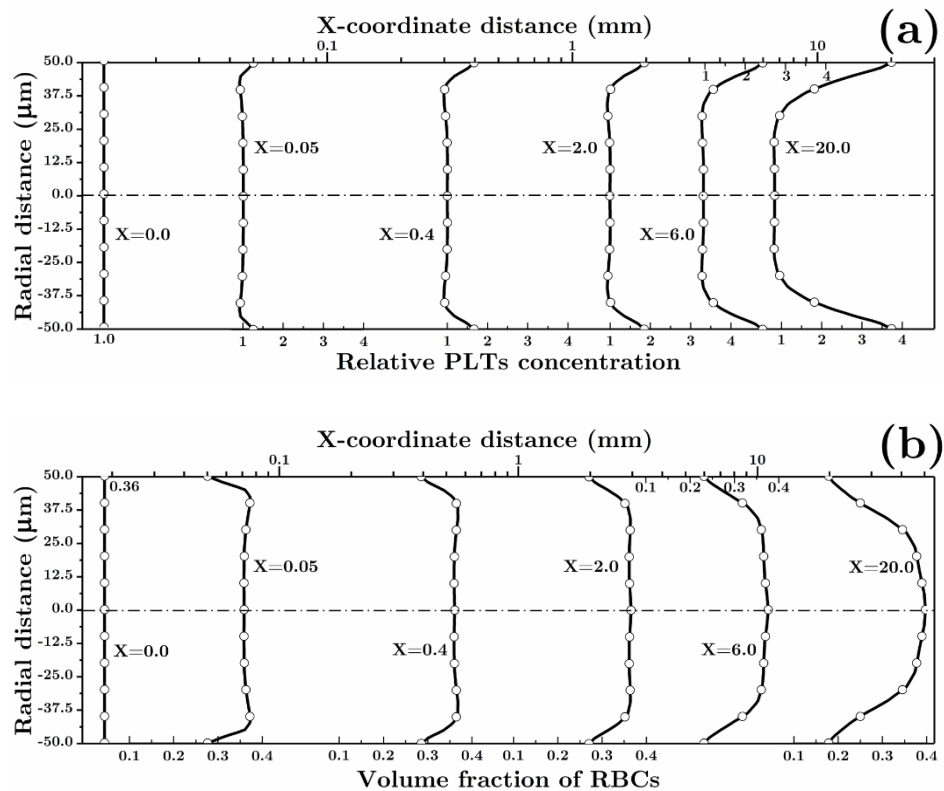

Figure 3 Axial development of the platelets (a) and the RBCs (b) distribution along the y-direction.

## 3.2. Flow in a sudden expansion micro-channel

Next, we consider flow of blood in a sudden expansion micro-channel. All the model parameters are the same as the previous case. Figure 4 shows the schematic of the sudden expansion micro-channel. The inlet velocity and the hematocrit are 0.413 m/s and 40% respectively (Rui Zhao et al., 2008). The platelets bulk concentration is $C_0 = 2.5 \times 10^{14} \text{PLTs/m}^3$ and the entrance length before the channel expansion is 20mm.

The comparison between the simulated results and the experimental observation for the relative platelets concentration along the y-direction is shown in Figure 5 (a), which shows a good agreement. The values based on the numerical simulations at a location 50μm after the expansion agree with the experimental data better than the simulated results at a location 20μm after the expansion. Figure 6 shows the platelets distribution by the simulation and the experiment, the RBCs concentration and the velocity (streamlines) field of the whole blood by the numerical simulation. Interestingly, after the sudden expansion, near the flow reattachment points, we notice a very thin layer where the platelets concentration is relatively low [see Figure 6 (a) for the two dimensional field and Figure 5 (b) for the quantitative plot.]. This pattern agrees with the observation of Karino and Goldsmith (Karino & Goldsmith, 1979a, 1979b). Their experiments showed that after the sudden expansion near the flow reattachment points a rapid decrease in the platelets deposition is achieved. The low platelets deposition may be due to the fact that near the flow separation/reattachment point, the RBCs concentration is relatively high [see Figure 6 (b)], and the platelets seemed to be pushed away from this region due to the RBCs-platelets collision.

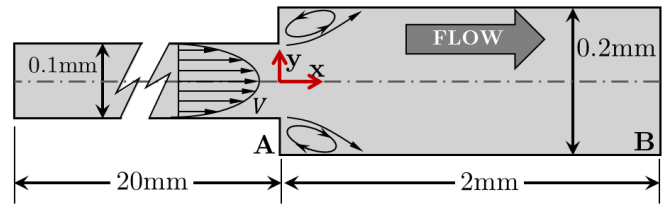

Figure 4 Schematic of the sudden expansion micro-channel. The inlet velocity is 0.413m/s. The hematocrit is 40%. The platelets bulk concentration is $C_0 = 2.5 \times 10^{14} \text{PLTs/m}^3$.

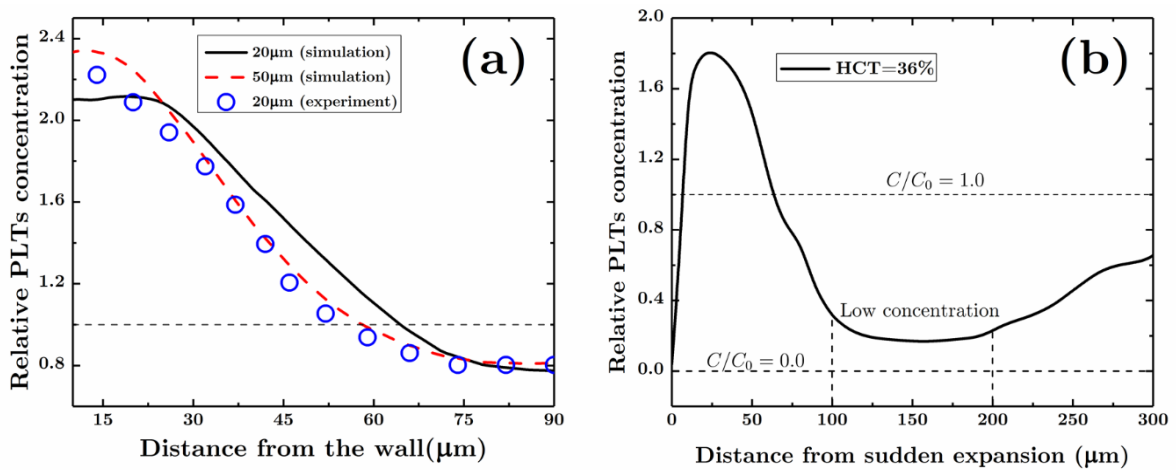

Figure 5 (a) Comparison of the numerical and the experimental results for the relative platelets concentration along the y-direction. (R. Zhao et al., 2010) (b) Platelet distribution along wall A-B after the sudden expansion. See Figure 4 for the definition of the wall A-B.

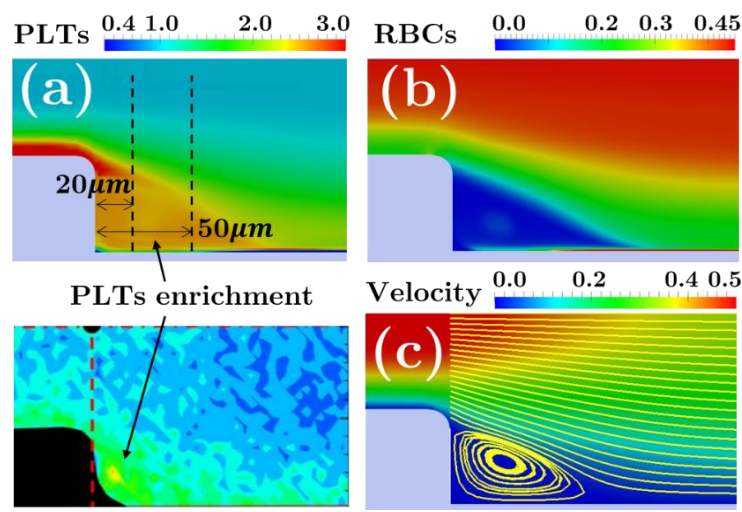

Figure 6 (a) The numerical (up) and experimental (down) relative platelet concentration field (Rui Zhao et al., 2008). (b) The numerical volume fraction fields of the RBCs. (c) Velocity field and streamlines of the whole blood; the unit of the scale bar is m/s. The results by simulations are at 1s. The experimental figures are reused by permission (Rui Zhao et al., 2008).

### 3.3. Flow in a rectangular micro-channel with crevices

Figure 7 shows the geometry of the case studied in this section. This particular application is motivated by a persistent problem in most blood- wetted devices, namely the seams and joints between component parts of a device or the flow circuit, which are predisposed to thrombus deposition. The inlet velocity and the hematocrit are 0.0238 m/s and 40%, respectively. The platelets bulk concentration is $C_0 = 3.0 \times 10^{14} \mathrm{PLTs/m^3}$ which was used in the thrombus growth (Wu et al., 2017). The experimental results are juxtaposed to microscopic images obtained using the method of Zhao et al. (2008) (Rui Zhao et al., 2008) in a microchannel of the same dimensions, and under similar conditions. Briefly, platelets-sized fluorescent particles (3 μm, Duke Scientific) were added to a suspension of RBC ghost cells (hematocrit = 40%) in a ratio of approximately 1:10. The sample solution is delivered by a syringe pump (Harvard PHD) at a prescribed flow rate, corresponding to the velocity in the simulations. The fluorescence signal is captured by a sensitive high-resolution CCD camera (SensiCam-QE, Cooke Corp).

From Figure 8, it can be seen that a good qualitative agreement for both the RBCs distribution and the platelets distribution is obtained. In particular, both results reveal a zone of low RBCs in the deeper regions of the narrow crevices and the slots [see Figure 8 (b)]. The high concentration of platelets in the RBCs deficit zone is shown in Figure 8 (c). This can be attributed to the exclusion of the platelets from the main flow by the RBCs. Figure 9 (a) shows the concentration fields for the platelets at different times. We see that after the flow has started, due to the presence of the RBCs, a high concentration of platelets move into the crevice quickly. Then as time passes, the platelets continue diffusing into the deeper parts of the crevice, but much slower since the effects of the RBCs induced diffusion is diminished in the deeper regions of the crevice. This platelets diffusion process is also indicated by Figure 9 (b) which shows the time evolution of the platelets accumulation in the crevice.

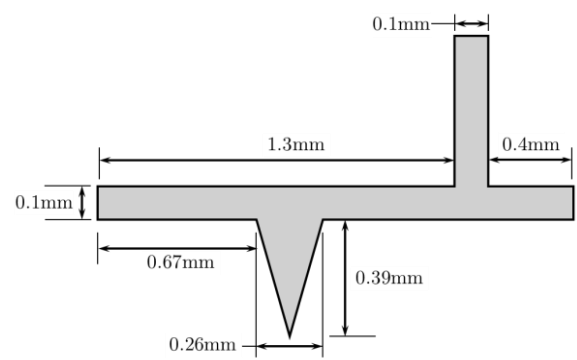

Figure 7 Schematic of the rectangular micro-channel with a depth of 0.07mm. The inlet velocity is 0.0238m/s. The hematocrit is 40%. The platelets bulk concentration is $C_0 = 3.0 \times 10^{14} \mathrm{PLTs/m^3}$.

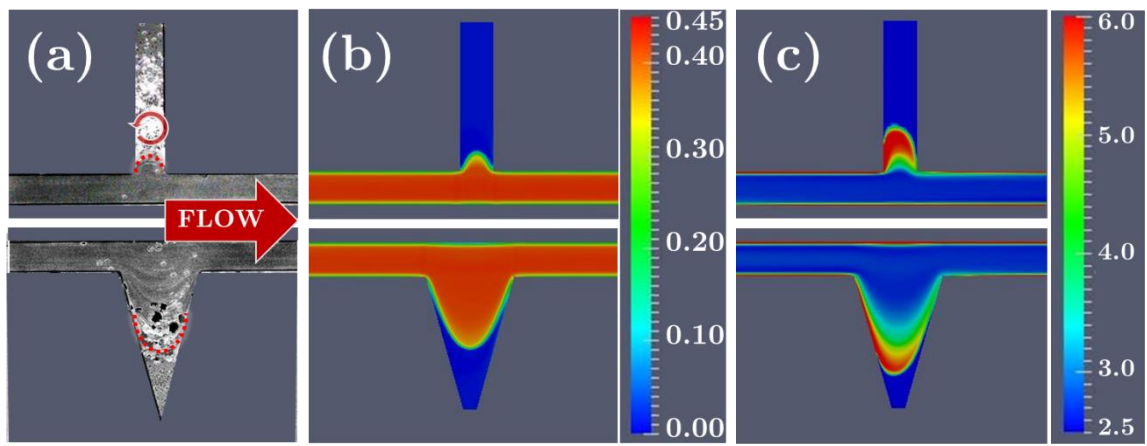

Figure 8 (a) Experimental observations of the RBCs ghosts seeded with 3μm (platelets sized) fluorescent particles. The bright regions correspond to the concentrated accumulation of platelets-sized fluorescent particles. The RBCs follow the main stream. (b) RBCs distribution predicted by Mixture Theory. (c) Platelets concentration predicted by the numerical simulation; the units of the scale bar is $1 \times 10^{14} PLTs/m^3$.

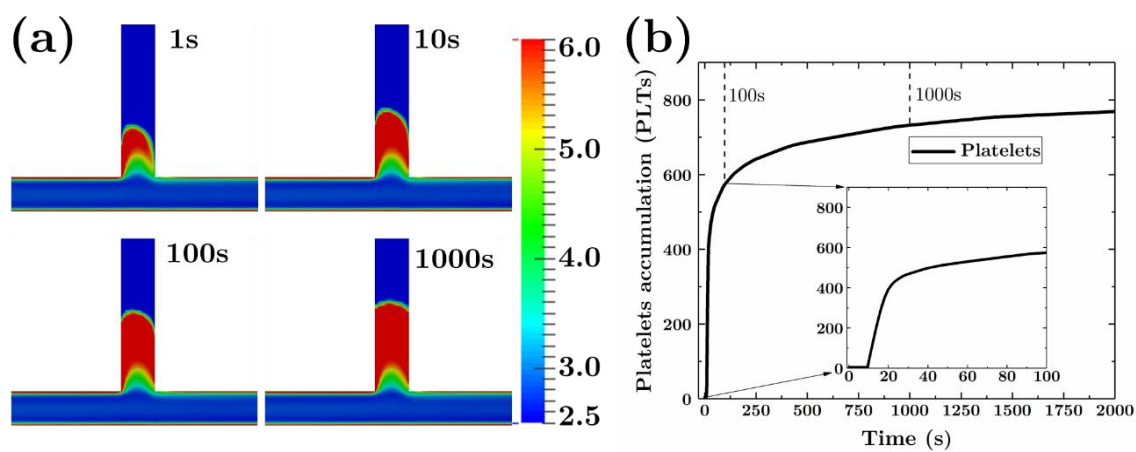

Figure 9 (a) The predicted instantaneous concentration fields of the platelets at different times. The units of the scale bar is $1 \times 10^{14} PLTs/m^3$. (b) Time evolution of the platelets accumulation in the crevice.

## 4. Discussions

The near-wall excess of platelets in a blood vessel is a well-known phenomenon, vital for the body's response to injury (Hund & Antaki, 2009). There are a few mathematical and computational methods used to model the inhomogeneous transport of the platelets: the mesoscale simulations which models the dynamics of each individual cells (Crowl & Fogelson, 2010; Dupin, Halliday, & Care, 2006; Vahidkhah, Diamond, & Bagchi, 2014; Wu, Martin, et al., 2016; Zhang, Johnson, & Popel, 2008), and the fluid-continuum models which treat platelets as a continuum (Jung & Hassanein, 2008; Jung, Lyczkowski, Panchal, & Hassanein, 2006; Wu et al., 2015). In this paper, we have proposed and formulated a new convection-diffusion equation to describe the platelets migration. With the proposed model, we have studied the non-uniform distribution of the platelets in several different benchmark problems, where the predicted concentration of the platelets can be about five times higher than the concentration near the centerline of the channel.

It is known that the enrichment of the platelets near the walls occurs when the Fahraeus-Lindquist effect is present, and this is usually the case in small arteries in the micro-scale range. The computational cost of the mesoscale simulations in micro-scale blood flow is not a critical element. Therefore during recent years, using mesoscale simulations, the near-wall enrichment of the platelets has been successfully predicted by numerous works (AlMomani et al., 2008; Crowl & Fogelson, 2010; Reasor et al., 2013; Skorczewski et al., 2013; Vahidkhah et al., 2014). However, in many problems, this micro-scale phenomenon may be coupled with the large/macro scale flow. For example, as demonstrated in Figure 7, in a millimeter-sized device there may be crevices that are in the micro-scale range; furthermore, in human body, the scale of the blood vessels varies dramatically. For these problems, computational cost of mesoscale simulations is prohibitive. However, using a continuum model developed here, such a computational difficulty can be conquered.

In the studies reported in the previous section, we notice that the platelets tend to concentrate in "dead zones". For the flow in a sudden expansion channel, this occurs in the circulation zone after the sudden expansion, while for the channel with crevices the platelets move to the deeper regions of the crevices. The high concentration of the platelets in these zones can increase the possibility of thrombosis (platelets deposition) and further causes serious problems in blood-wetted devices. The shear stress of the (whole) blood in the circulation zone is reasonably small which implies that the accumulated platelets can hardly be moved by blood flow in these zones. The residence time of the platelets entering these zones are much longer compared to the platelets, which are in the main stream; therefore the platelets in this circulation zone have much longer time and larger possibility to be activated, aggregate together and deposit on the channel walls. Furthermore, the deposited platelets in the "dead zones" is possibly able to activate and capture the upstream (the incoming) platelets, which indicates that these "dead zones" can be nidus for serious device thrombosis (Jamiolkowski, Pedersen, Wu, Antaki, & Wagner, 2016; Wu et al., 2017).

## 5. Conclusions

Motivated by the observed phenomenon of platelets enrichment near the walls, we have used a multicomponent continuum theory to model the blood flow. We have also proposed and tested a mathematical model for the platelets transport, which includes the effect of blood cells collisions. The model accurately predicts the spatially non-uniform distribution of the platelets in benchmark problems, agreeing with the experimental observations both quantitatively and qualitatively. The numerical simulations indicate that in the rectangular channel near the walls, the concentration of the platelets can be about five times higher than the concentration near the channel centerline. Our simulations also reveal that the platelets tend to accumulate in the recirculation zones and in the micro-crevices of a channel; these regions may become nidus for excessive thrombosis. Through successive numerical study on the above benchmark problems, we believe that the model is useful for studying practical problems such as thrombosis in practical medical devices or in patient specific pathological vessels. We should mention that the model developed and used here is only appropriate for a healthy human, and it does not capture any blood disorder. Even though we have ignored the biochemical effects of blood in this paper, the model presented here can be extended to study such cases [see Anand et al (2006) (M. Anand et al., 2006a)].

## 6.    Acknowledgments

This research was supported by NIH grant 1 R01 HL089456.